\documentclass[twocolumn,showpacs,preprintnumbers,amsmath,amssymb]{revtex4}
\usepackage{graphicx}
\usepackage{dcolumn}
\usepackage{bm}
\begin{document}
\title{
        On the magnetic dipole energy expression of an arbitrary current distribution
\\}
\author{Keeyung Lee \\
}
\address{
Department of Physics, Inha University,
Incheon, 402-751, Korea\\
}
\date{\today}
\begin{abstract}
     We show that the magnetic dipole energy term appearing in the expansion of the magnetic potential energy of a localized current distribution has the form $ U= + \bf{m} \cdot \bf{B}$ which is wrong by a sign from the well known $ - \bf{m} \cdot \bf{B}$ expression. Implication of this result in relation to the electric dipole energy $ - \bf{p} \cdot \bf{E}$, and the force and torque on the magnetic dipole based on this energy expression is also discussed.
\end{abstract}
\narrowtext
\maketitle

\vfil\eject
{\bf I. Introduction}

Potential energy of a charge distribution in an external electric field or the potential energy of a current distribution in an external magnetic field is an important concept in the electromagnetic theory. It is well known that in the energy expansion of a localized charge or of a localized current distribution, the dipole energy term becomes important.

For an electrically neutral system, since the monopole term vanishes, electric dipole energy which has the form $ U= - \bf{p} \cdot \bf{E}$ becomes the most important term in most cases. One elementary method of deriving this expression is to consider the potential energy of two point charges $\pm q$ in an electric field due to the electrostatic potential $\phi ( \bf{r} )$. In that case, the potential energy can be written as $ U= +q {\phi(\bf{r} +\bf{d})} +(-q) \phi(\bf{r})$, which gives the potential energy expression  $ - \bf{p} \cdot \bf{E}$ by taking the first order approximation $ \phi (\bf {r} + \bf{d} ) = \bf{d} \cdot \nabla \phi (\bf {r}) $. This expression can also be derived by calculating the amount of work needed to rotate the dipole, if the torque expression $\tau = pE \sin \theta $ has already been considered.

It is also well known that the magnetic dipole energy has the form $ U= - \bf{m} \cdot \bf{B}$, when the dipole is located inside an external magnetic field $\bf{B}$. Elementary derivation of this expression can be done, first by calculating the amount of work needed to rotate the dipole, if the torque expression on the dipole $\tau = mB \sin \theta $ has already been considered.
However, compared with the electric dipole case, derivation of such torque expression is a relatively difficult task. The derivation of torque expression on a system with an arbitrary geometry is not straightforward, and usually a square shaped loop current model is assumed\cite{hall,feyn}.
Some textbooks directly consider the force on the rectangular or on the circular loop instead of deriving the torque expression\cite{purc,grif}. Magnetic dipole energy expression is then obtained by using the calculated force expression in the energy-force relation  $ \bold{ F} = - \nabla {U} $.
Many advanced level textbooks also adopt such indirect derivation approach, but with systems generalized to arbitrary geometry\cite{jack1,land}. However, derivation of the force expression of an arbitrary current distribution itself is not an easy task at all\cite{boy}.

In this work, we consider direct expansion of the magnetic potential energy and show that the expansion of the magnetic energy of an arbitrary current distribution system results in the expression $ + \bf{m} \cdot \bf{B}$ for the dipole energy term. Implication of this result in connection with the force and torque on the dipole is also discussed.

\bigskip
{\bf II. Magnetic Energy Expansion}

Before discussing the magnetic energy case, let us consider first the expansion of electric potential energy $ U = \int dV \rho(\bf{r}) \phi(\bf{r}) $ of a static charge distribution with charge density $\rho (\bf {r} )$ under the influence of an external potential $\phi( \bf{r})$.
It is straightforward to show that this energy expression can be expanded as

 $$ U = q \phi (0) - {\bf{p}} \cdot {\bf{E}}(0) + ...  \eqno(1)$$
in which $q$ is the amount of total charge, $\bf{p}$ is the dipole moment, and ${\bf{E}} (0)$ is the electric field at an arbitrary point inside the charge distribution. Therefore we find that the electric dipole energy of a sufficiently localized charge distribution system can be approximated as $ - \bf{p} \cdot \bf {E}$

Now let us consider the magnetic energy of a current distribution with current density $\bf{J}$ under the influence of an external magnetic field. When the vector potential which represents the magnetic field is $\bf{A (\bf{r})} $, the magnetic potential energy can be expressed as,

$$ U= \int dV \mathbf{J}\cdot \mathbf{A}
 = \sum_{k=1}^3 \int dV J_{k}  A_{k}  \eqno(2)$$
Now, for a sufficiently localized current distribution, this energy expression can be expanded as follows,

 $$ A_{k} (\mathbf{r}) = A_{k} (0) + \mathbf{r} \cdot [\nabla A_{k}]_{0} + \frac{1}{2} \sum _{i=1}^3\sum_{j=1}^3 x_{i} x_{j} [\frac {\partial^2 A_{k}}  {\partial x_{i} \partial x_{j}} ]_0 +...  \eqno(3)$$
where the origin of the coordinate is chosen inside the current distribution.
Using such expansion, the magnetic energy can be expressed as $U = U_0 + U_1 + ...$, where

 $$U_0 = \sum_{k=1}^3 \int dV J_{k} A_{k}  = \sum _{k=1}^3 A_{k} \int dV J_{k}  \eqno(4)$$
 and

 $$U_1 = \sum _{k=1}^3 \int dV J_{k} \mathbf{r} \cdot [\nabla A_{k} ]_0  \eqno(5)$$
which are the monopole and dipole terms respectively. For a stationary current system, the monopole term $U_0$ vanishes since  $\int J_{k} dV = 0 $.
Now, if the constant vector $[\nabla A_{k} ]_0 $ is temporarily denoted as $\mathbf{c}$, the dipole term $U_1 $ can be expressed as,

$$U_1 =  \sum  _{k=1}^3 U_{1k} \eqno(6)$$
where $U_{1k} =  \int dV J_{k}(\mathbf{r} \cdot\mathbf{c})$. 
$U_{1k}$ term can now be expressed as, 
   
$$ U_{1k} =  \int dV \ x_{k} ( \mathbf{J} \cdot \mathbf{c})   - \int dV [ \mathbf{c}\times ( \mathbf{r} \times \mathbf{J })]_{k}  \eqno(7)$$
if the vector identity $ \mathbf{c} \times ( \mathbf{r} \times \mathbf{J} ) = \mathbf{r} (\mathbf{J} \cdot \mathbf{c}) - \mathbf{J} (\mathbf{r} \cdot \mathbf{c} )$ is used. 
It could be noted here that the integral of the second term on the right hand side can be expressed as $ 2 [\mathbf{c} \times \mathbf{m}]_{k} $ where $\bf{m}$ is the magnetic dipole moment which is defined as $ \mathbf{m} = \frac{1}{2} \int \mathbf{r} \times \mathbf{J} dV $.

Let us now consider the first term on the right hand side of Eq.(7). Since $\bf{c}$ is a just constant vector, this term can be written as $\sum_i c_i \int x_{k} J_i$. Therefore, using the relation $ \int dV (x_{k} J_i + x_i J_{k} ) =0 $ which is satisfied for a stationary current system\cite{note1}, this term can now be expressed as $-\sum_i c_i \int dV x_i J_{k}$, or, as $-\int dV J_{k}(\bf{r}\cdot\bf{c}) $, which means that the first term reduces to $-U_{1k}$. Therefore we get the expression,

$$ U_{1k}  =  -[\bf{c} \times \mathbf{m}] _k \eqno(9)$$
in which $\bf{c}$ stands for $[\nabla A_k ]_0 $.

Therefore, using the Levi-Civita symbol and Einstein summation convention, we finally obtain the expression for the magnetic dipole energy as $ U_1 = - \varepsilon_{kij} ( \partial_i A_k ) m_j $, or 

$$ U_1 =  \varepsilon_{jik} ( \partial_i A_k ) m_j \eqno(10)$$
using the $\varepsilon_{kij} = -\varepsilon_{jik}$ property .
Now since $ \varepsilon_{jik} \partial_i A_k = B_j $ from the relation $\bf{B} = \nabla \times \bf{A}$, we find that the magnetic dipole energy can be written as $ + \mathbf{m} \cdot \mathbf{B} (0)$, in which $\bf{B}(0)$ is the magnetic field at an arbitrary point inside the current distribution.  We may note that this dipole term is the only term left in the expansion for the particular case of uniform external magnetic field. To show this point, let us note that the vector potential of the form $\bf{A} = \frac{1}{2} \bf{r} \times \bf{B} $ produces a uniform magnetic field.
Since the vector potential in this expression is of the first order in the coordinate $x_i$, it is obvious that $[{{\partial^2 A_{k}} \over {\partial x_{i} \partial x_{j}}}]_0 $ is equal to 0, which means that the quadrupole term, as well as all remaining higher order terms vanish.

We have shown that the dipole term in the expansion of the magnetic potential energy is expressed as $ + \mathbf{m} \cdot \mathbf{B}$, whereas the dipole term in the electric energy expansion has the $ -\mathbf{p} \cdot \mathbf{E}$ form, which is an unexpected result considering that the electric dipole and the magnetic dipole have so many similar properties.
Although this result is rather unfamiliar, such 'wrong' sign of the magnetic dipole energy has already been discussed by D. J. Griffiths \cite{grif}. In that paper, the author derives the expression $ U = + \mathbf{m} \cdot \mathbf{B}$ using the the magnetic charge model and provides a rather lengthy explanation on why such opposite sign result should be obtained.

The fact that the energy of a magnetic dipole under the influence of an external field should be expressed as $ +\mathbf{m} \cdot \mathbf{B}$ could be confusing to many students, since the energy of a magnetic dipole is usually given as $ -\mathbf{m} \cdot \mathbf{B}$ in most textbooks\cite{hall,jack1}.
Such sign difference in magnetic dipole energy does not make any difference when the magnitude of force on the magnetic dipole only is of interest. But it results in the opposite direction of the force or torque on the dipole, and a good understanding on this situation may be important pedagogically.

\bigskip
{\bf III. Discussion on the Force and Torque on the Magnetic Dipole }

Before discussing the force and torque on the dipole, let us make sure that $ + \mathbf{m} \cdot \mathbf{B}$ is the right expression of the magnetic dipole energy. To show that point, consider situations where a current loop is placed in an uniform external magnetic field, in which magnetic dipole is directed (1)parallel (2)anti-parallel and (3)perpendicular to the magnetic field direction.
Considering the superposed field of the external field and of the field due to the current loop and calculating the  magnetic field energy $ \int dV |\bf{B} |^2  $, it is obvious that the energy is increased for case (1) and is decreased for case (2). It is also obvious that there is no change in the magnetic energy for case (3). This example shows that the energy variation for each situation is consistent with the magnetic energy expression $ U = + \mathbf{m} \cdot \mathbf{B}$.

Now let us consider the force and torque on the magnetic dipole. If one attempt to apply the $ \bold{ F} = - \nabla {U} $ relation between the force and potential energy using the energy expression $ U = + \mathbf{m} \cdot \mathbf{B}$, the force on the magnetic dipole becomes $ \bf{F} = - \nabla ( \bf{m} \cdot \bf{B} )$ which is wrong by a sign from the right expression\cite{jack1}. We again do not get the right expression for the torque if $ U = + \mathbf{m} \cdot \mathbf{B}$ is used in connection with the common expression $\tau (\theta) = - dU/{d \theta }$, since we get the result $\tau (\theta) =  +mB \sin \theta$, which shows that the torque is exerted in the wrong direction, i.e. the dipole tends to orient anti-parallel to the magnetic field. 

This situation seems to indicate that the expression $ U = + \mathbf{m} \cdot \mathbf{B}$ is wrong by a sign, and suggests that the right force expression $ \nabla ( \bf{ m} \cdot \bf{B})$ is obtained if the expression $ U = - \mathbf{m} \cdot \mathbf{B}$ is is used. In fact, the exact reverse steps are followed in the textbook by J.D. Jackson\cite{jack1}, in which the author derives the dipole energy expression $ U = - \mathbf{m} \cdot \mathbf{B}$ using the calculated force expression, but with a careful remark that this is the case for the "permanent magnetic dipole" only\cite{jack1}. In that textbook, the author also provides a comment that "$ U = - \mathbf{m} \cdot \mathbf{B}$ is not the total energy of the magnetic moment in the external field", which is intended to emphasize that there could be an exchange of energy between the dipole and the external source.
This implies that if the magnetic moment does not have a fixed magnitude as for the case of spin magnetic moment, then exchange of energy could occur between the magnetic dipole and the external source. Such situation is nicely discussed by J.R. Reitz {\it et al.}\cite{reitz}. In that textbook, a rigid circuits system is used to explain why the force and magnetic energy relation should be expressed as  $ \bold{ F} =  \nabla {U} $ rather than  $ \bold{ F} = - \nabla {U} $, when a current loop is placed under the influence of external field due to all other circuits. Using that argument, it could be shown that the energy supplied by the external source to a circuit loop is exactly twice the potential energy change of the circuit\cite{reitz,note2}.
The mechanism of energy exchange between a current loop and all others in this situation is found to be the electromagnetic induction. For example, if a displacement is made on a current loop, a change of magnetic flux in the loop is accompanied and induced electromagnetic effect arises. This means there will be an exchange of energy  between the current loop and all others. Of course, no such exchange of energy occurs between a permanent magnetic dipole and the external magnetic field, or between an electric dipole and the external electric field.

\bigskip
{\bf IV. Acknowledgement }
This work was supported by the Inha University Research fund.


\begin{references}
\bibitem{note1} This relation can be proved by employing the vector identity $ \nabla \cdot( x_i x_j \mathbf{J} ) = J_i x_j + J_j x_i $.
\bibitem{note2} It is interesting to note that the exactly similar situation occurs in the electrostatic system, in which a dielectric slab is inserted inside a parallel plate capacitor connected to a battery. In that case, it could be shown that the energy supplied by the battery is exactly twice the magnitude of the change of the capacitor stored energy.
\end{references}
\end{document}